\documentclass[10pt,journal,compsoc]{IEEEtran}

\ifCLASSOPTIONcompsoc
\usepackage[nocompress]{cite}
\else
\usepackage{cite}
\fi

\ifCLASSINFOpdf
\usepackage[pdftex]{graphicx}
\graphicspath{{images/newPlots/}{images/}}
\DeclareGraphicsExtensions{.pdf,.jpg,.jpeg,.png}
\else
\usepackage[dvips]{graphicx}
\graphicspath{{images/newPlots/}{images/}}
\DeclareGraphicsExtensions{.eps}
\fi

\usepackage[cmex10]{amsmath}

\usepackage{array}
\usepackage{booktabs}

\usepackage{fixltx2e}
\usepackage{placeins}
\usepackage{rotating}

\usepackage[hyphens]{url}
\usepackage[hidelinks]{hyperref}

\usepackage[capitalise]{cleveref}

\crefname{equation}{}{}

\newlength{\mywidthA}
\newlength{\mywidthB}
\newlength{\mywidthC}
\begin{document}
\bstctlcite{IEEEexample:BSTcontrol}

\title{Serving the Grid: an Experimental Study of Server Clusters as Real-Time Demand Response Resources}


\author{Josiah~McClurg,~\IEEEmembership{Student Member,~IEEE,}
	Raghuraman~Mudumbai,~\IEEEmembership{Member,~IEEE}
	\IEEEcompsocitemizethanks{\IEEEcompsocthanksitem Authors are with Department
		of Electrical and Computer Engineering, University of Iowa, Iowa City,
		IA, 52242 USA.\protect\\
		E-mail: [josiah-mcclurg OR raghuraman-mudumbai]@uiowa.edu}
	\thanks{Manuscript prepared fall, 2016}}

\IEEEtitleabstractindextext{%
	\begin{abstract}
		Demand response is a crucial technology to allow large-scale penetration of intermittent renewable energy sources in the electric grid. This paper is based on the thesis that datacenters represent especially attractive candidates for providing flexible, real-time demand response services to the grid; they are capable of finely-controllable power consumption, fast power ramp-rates, and large dynamic range. This paper makes two main contributions: (a) it provides detailed experimental evidence justifying this thesis, and (b) it presents a comparative investigation of three candidate software interfaces for power control within the servers. All of these results are based on a series of experiments involving real-time power measurements on a lab-scale server cluster. This cluster was specially instrumented for accurate and fast power measurements on a time-scale of 100 ms or less. Our results provide preliminary evidence for the feasibility of large scale demand response using datacenters, and motivates future work on exploiting this capability.
	\end{abstract}
	
	}

	\maketitle

	\IEEEdisplaynontitleabstractindextext

	%
	\IEEEpeerreviewmaketitle

	\IEEEraisesectionheading{\section{Introduction}\label{sec:intro}}


\IEEEPARstart{T}{his} paper uses experimental results from a power-aware
cluster to show how the fast response time, fine-grained controllability, and
large dynamic range of computer servers can allow 
computing clusters to provide flexible, large-scale, real-time
demand response services to the grid. Furthermore, as a step toward
implementing these services on a larger scale, we present an empirical
comparison of several candidate power-modulating interfaces available to modern
servers.

Datacenters have recently become an important and increasing portion of the
electric load in the US electric grid \cite{koomey2011growth}. However, in
their current configuration, datacenters are not particularly friendly to the
grid. The largest of them can require hundreds of megawatts guaranteed capacity
\cite{largestDatacenters}, but frequently experience large, unpredictable
fluctuations in actual power consumption \cite{powerFluctuations}.

At the same time, datacenters also have a significant amount of {\it excess capacity}.
In other words, there is considerable evidence (summarized below) that shows
that datacenters operate at substantially less than peak utilization most of
the time. This suggests the possibility that the ``spare capacity" of computer servers
can be used to provide load-shaping services to the electric grid and
indeed this idea has been proposed \cite{2014DemandResponseSurvey}, and is being
actively studied by researchers
\cite{2012DatacenterDRFieldStudies,2013DatacenterCLR,2013RealtimePowerControlOfDatacenters,2014DatacenterDemandResponse,2014DatacenterDynamicPricing,2014DatacenterGridLoadStabilizer,2014DatacentersAndElectricVehicles,2014PricingDatacenterDemandResponse,2015DatacenterOptimalRegulationService}.

This paper represents a new contribution to this literature by specifically
considering {\it real-time} demand response at much shorter time-scales than
previous work in this area. We argue that compared to other types of electric
loads, computer servers are especially well-suited to provide real-time
services because their power consumption is controllable to (a) a fine
granularity over (b) a large dynamic range of power levels with (c) fast ramp
rates, and we back up these arguments with detailed experimental evidence.


\subsection{The importance of load-shaping}

The idea of using load shaping to increase power grid reliability has been
around since Edison's time \cite{1915MultiplexCostAndRateSystem}. However, grid
operators have traditionally not emphasized load-side management, relying
instead on the paradigm of providing energy on demand to completely {\it
passive} consumers who have essentially unrestricted ability to vary their
usage over time.

Indeed, active load-side management was arguably redundant in the traditional
power grid. Grid operators have long been aware of the high degree of
statistical regularity in 
electricity demand over time on
daily, weekly and yearly time-scales and were traditionally able to take
advantage of these patterns to predict loads and optimize generation schedules
accordingly \cite{1987ShortTermLoadForecasting}.

This situation has changed dramatically with the increasing penetration of
intermittent renewables like wind and solar in the grid; renewable energy
generation has proved to be far less predictable than load
\cite{porter2012survey}, and treating renewables as ``negative loads'' is not
only expensive and wasteful \cite{2013PowerSystemSchedulingWithRenewables} in
many ways, it also strains the existing reliability and stability mechanisms of
the grid \cite{2007IntermittentGeneration}. This has led to a growing
recognition that large-scale energy storage and/or ``virtual storage''
\cite{2011DemandResponseSmartLoads,2014DemandResponseSurvey} in the form of
demand response are essential to accommodate renewable energy sources in the
grid.

\begin{figure}
	\centering
	\includegraphics[width=\mywidthA]{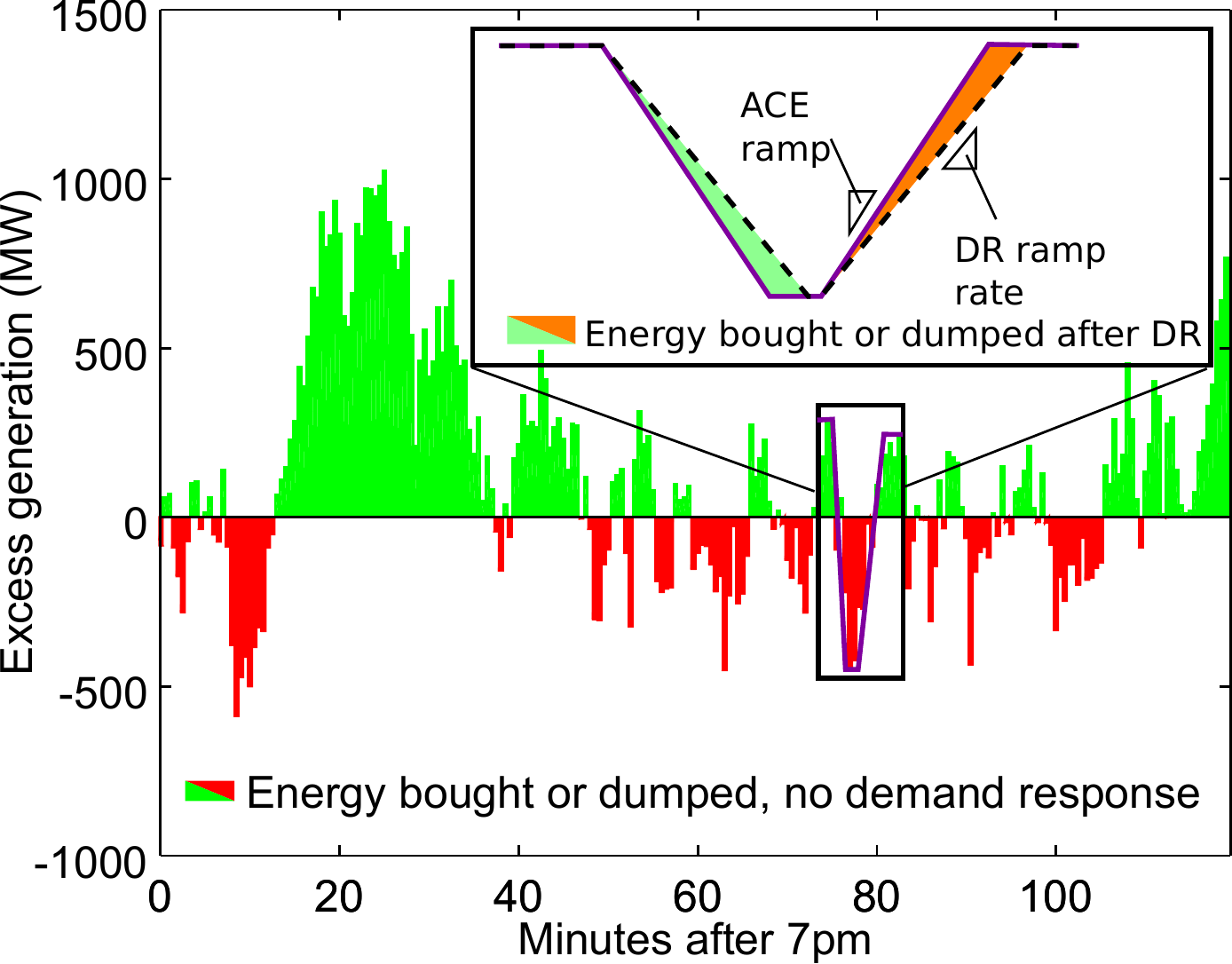}
	\caption{MISO area control error}\label{fig:demandResponseAndPowerDip}
\end{figure}

A simple case study illustrates how realtime demand response can provide a benefit over slower demand response. \cref{fig:demandResponseAndPowerDip}, plots the area control error (ACE) of the Midcontinent Independent System Operator (MISO) over a two hour period \cite{misoACE}. During this period, $460 \text{ MWh}$ of energy is dumped and $200 \text{ MWh}$ of unscheduled energy purchases are made. Simulations show that using $200 \text{ MW}$ demand response capacity with a ramp rate of $10 \text{ MW}/\text{min}$ reduces the dumped energy by a mere $70 \text{ MWh}$. Increasing the ramp rate to $100 \text{ MW}/\text{min}$ reduces the dumped energy by $190 \text{ MWh}$. And, similar results hold for the purchased energy reductions.

\subsection{Real-time demand response}

It is useful for our purposes to classify demand response techniques according
to how far in advance the load adjustments need to be determined, and how
smooth these adjustments must be. Specifically, we make a distinction between 
{\it real-time} and {\it offline} demand response methods based on whether the 
time-scale for load adjustments is significantly shorter or longer than 
a threshold which we take to be $30$ seconds.

This choice of threshold in the definition of a {\it real-time} demand response
technique is not completely arbitrary. International standards require the time
constants of the primary control of generators participating in the frequency
regulation process to be on the order of a few tens of seconds
\cite{primarycontrol}. This in turn determines the {\it fastest} time-scale on
which it is possible to perform load shaping without risking
instability\footnote{At short time-scales, another interesting possibility is to have adjustable loads actively participate in the frequency regulation	function itself; a detailed exploration of this idea is beyond the scope of this paper, but there exists some previous work e.g. \cite{2014DatacentersAndElectricVehicles} in this area which remains an active topic of research.}.

Examples of electric loads that are unsuitable for real-time demand response
include residential lighting and HVAC systems and also industrial loads
consisting of large rotating machines. In contrast, computer servers and
electric vehicle charging systems \cite{2014DatacentersAndElectricVehicles} are
examples of loads whose power consumption is not limited by the inertia of
moving mechanical components and are good candidates for real-time demand
response.

\subsection{Managing power consumption in datacenters}

Ever since large-scale datacenters have existed, power considerations have been
recognized as economically crucial to their operation and operators have
developed sophisticated methods
\cite{2011GreenScheduling,2013ParasolAndGreenSwitch,energyAwareResourceAllocation} for managing and
controlling the two major categories of power consumption in datacenters: (a)
the building maintenance and cooling systems, and (b) the computer servers
themselves. As discussed earlier, we focus in this paper on the latter i.e. the
power consumption in the servers. We note that designing datacenters to be
energy-efficient and {\it minimizing} their power consumption \cite{2014GreenEnergyAwarePowerManagement}
is quite different from providing load-shaping services to the grid.

An empirical study by Lawrence Berkeley National Laboratory (LBNL) discussed
the advantages and disadvantages of various datacenter demand response
strategies \cite{2012DatacenterDRFieldStudies}. One power control method widely
studied in the literature is the geographical migration of virtual machines
\cite{WangCLR2013} or transactional workloads and batch jobs
\cite{ChengWorkload2014} in response to variations in hour-ahead locational
marginal pricing.

There is also a significant literature on determining optimal pricing
strategies for datacenters to offer load-shaping serves on electricity markets
see e.g. \cite{2013DatacenterCLR, 2014DatacenterDynamicPricing,
2014PricingDatacenterDemandResponse}.

Note, however, that price bids on electricity markets are settled and refined
at relatively long time-scales, ranging from day-ahead and hour-ahead
to near ``real-time" markets which operate $5$ minutes in advance. Power
fluctuations at timescales shorter than $5$ minutes must be settled by the
system operator (ISO) through explicit signaling mechanisms. Typically, the ISO
monitors the Area Control Error \cite{primarycontrol} measurements of inter-area
power-flow imbalances and uses these measurements to generate regulation
service requests \cite{PJMbalancingOps} periodically on the order of every
$10$ seconds or so. Through these requests, the ISO orders providers to change
their active power consumption by specific amounts. Typically, these providers are
compensated in advance by the ISO for providing such services on demand either
through market bids, or long-term contractual arrangements \cite{PJMmanual28}.

In this paper, we consider a model where a datacenter has been contracted to
provide {\it real-time best effort regulation service} to the grid. 
Accordingly, we assume that a grid operator periodically sends service requests
in the form of desired active power operating points to the datacenter operator
who controls the servers to meet these requests as long as the requests are
feasible given the requirements of the server workloads.

We deliberately chose this ``best effort" service model to eliminate by
assumption the possibility of any conflict between the grid's service requests
and the quality of service (QoS) provided to the datecenter's software
workloads. More sophisticated models to provide maximum grid regulation service
while maintaining statistical or deterministic QoS are left to future work (an
interesting first step along those lines is
\cite{2013RealtimePowerControlOfDatacenters}). We focus narrowly in this
paper on a detailed empirical study of different software mechanisms to meet
grid service requests that are presumed to be feasible. 


\subsection{Organization}

The rest of this paper is organized as follows. \cref{sec:proofOfConcept} shows that it is possible to achieve good power tracking performance with a simple distributed controller. \cref{sec:suitability} explores the demand response power metrics of ramp rate, dynamic range, and tracking error. \cref{sec:interfaces} examines detailed power statistics of several different cpu-capping software interfaces. 
\cref{sec:concl} is the conclusion and suggests future work.

\begin{figure*}
	\centering
	\includegraphics[width=\mywidthC]{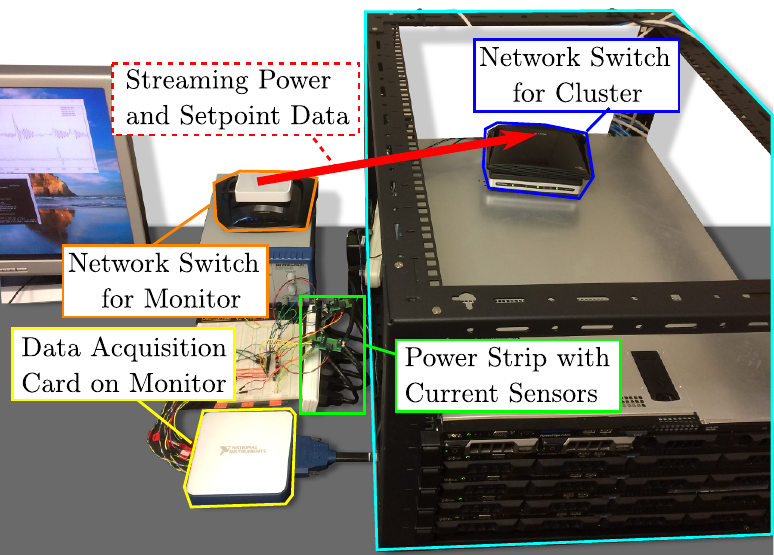}
	\caption{Experimental setup}\label{fig:serverRack}
\end{figure*}
\section{A Simple Power-Tracking Experiment}\label{sec:proofOfConcept}

Our first experiment is intended to illustrate that very simple software methods can be effective in accurately controlling the real-time power consumption of servers under favorable conditions.

First, however, we describe the hardware and software setup of the server cluster on which all the results in this paper are based. The cluster is pictured in \cref{fig:serverRack} and was chosen as a small-scale model that approximates as closely as possible commonly-used server hardware configurations in datacenters \cite{koomey2007servers,koomey2011growth}.

The cluster consists of four standard Dell PowerEdge R320 rack servers each powered by an Intel Xeon E5-2400 series processor with 6 cores and hyperthreading enabled. The servers all run 64-bit Ubuntu Server operating system version 15.04 with the 3.19.0-43-generic Linux kernel, and power management options are set to their default settings.

To obtain faster power measurements than are available from a commercial power distribution unit (PDU), we powered the servers through a specially instrumented power strip. A National Instruments PCIe-6323 data acquisition card measures the wall voltage and the output of amplified current shunts (Linear Technology LT1999 amplifier and a $0.02 \; \Omega$ resistor, with combined nominal current measurement tolerance of $\pm 1.6 \%$). A standard desktop PC serves as a monitor computer that reads five channels (one for rack voltage, and four for individual server currents) at $10 \text{ kS/s}$ from the data acquisition card.

On the monitor computer, a software application polls rack-level power measurements from the data acquisition card, averages these measurements over $100 \text{ ms}$ windows\footnote{We did not perform zero-crossing detection when choosing the borders for the block averaging operation on the instantaneous power. So, the measurements presented throughout this paper have a higher variance than would have been obtained from averaging real power over an integer number of cycles. In other words, the tracking error measurements are somewhat conservative values.}, subtracts this value from the target power $P_{set}(t)$, and streams samples of the power tracking error\footnote{We actually stream the power measurements at $10 \text{ S/s}$, stream the power target only when it changes, and calculate the tracking error separately on each server. The result is functionally equivalent to streaming $e(t)$ at $10 \text{ S/s}$.} $e(t)$ over Ethernet UDP multicast packets at a rate of $10 \text{ S/s}$. $P_{set}$ represents a desired total power setpoint for all the servers in the cluster combined; in a larger-scale version of our cluster, $P_{set}$ can be thought of as the signal that the electric grid sends to request demand response services from the cluster. In all the experiments described in this paper, the servers independently perform demand response using only the total power tracking error signal $e(t)$ that is common to all servers.

\begin{figure}
	\centering
	\includegraphics[width=\mywidthA]{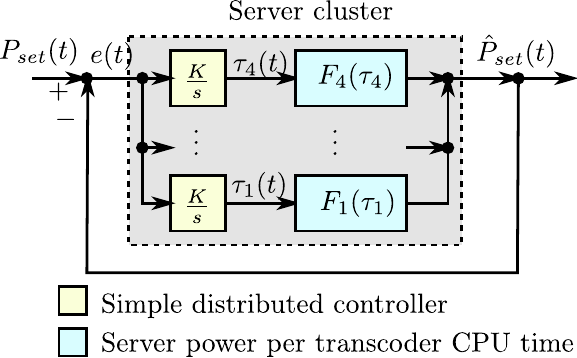}
	\caption{Block diagram of simple control scheme}\label{fig:controlScheme}
\end{figure}

We are now ready to describe our power-tracking experiment. In this experiment, the cluster is programmed to use the Linux avconv program 
to transcode video chunks from an NFS-mounted data store in a manner similar to the dynamic GOP transcoding scheme described in \cite{2004VideoCodingClustersOfWorkstations}. Each server has a worker which grabs the next available $10$-minute block from the shared video source file, transcodes that block, and writes the result back to the shared drive, for later concatenation. The source files are chosen to be large enough that the transcoding takes longer than the duration of the power tracking test. We chose the highly parallelizable transcoding application because it is especially well-suited for power tracking using simple controllers running independently on each server. It is worth noting that some major commercial transcoding services (such as Amazon Elastic Transcoder \cite{amazonTranscoderFAQ}) offer pay-per-video pricing at a best-effort conversion speed that is similar to the application we use in this experiment. Furthermore, by replacing the simple controller in \cref{fig:controlScheme} with the one described in \cite{joeThesis}, we can achieve realtime power tracking with soft service level agreement (SLA) enforcement. But, that is outside of the scope of this paper.

\subsection{Power Tracking Controller}
We assume that our realtime controllable load cluster has a demand response regulation service agreement in the PJM market. According to \cite{PJMbalancingOps,PJMmanual28}, the cluster must respond to a signal $s(t) \in [0,1]$ from the utility, so that its measured power consumption (measured in $10$-second intervals) is close to $P_{set}$ in \cref{eq:powerModel} with $D$ and $B$ being the dynamic range and base load agreed upon in the contract.

\begin{equation}\label{eq:powerModel}
P_{set}(t) = D s(t) + B
\end{equation}

In this experiment, we choose that $B$ to be cluster's power consumption with all the servers at idle, and $D$ to be the maximum possible dynamic range. We choose $s(t)$ to vary in a piecewise-constant manner over time to show the cluster's closed-loop response to a step input.

\begin{figure}
	\centering
	\includegraphics[width=\mywidthA]{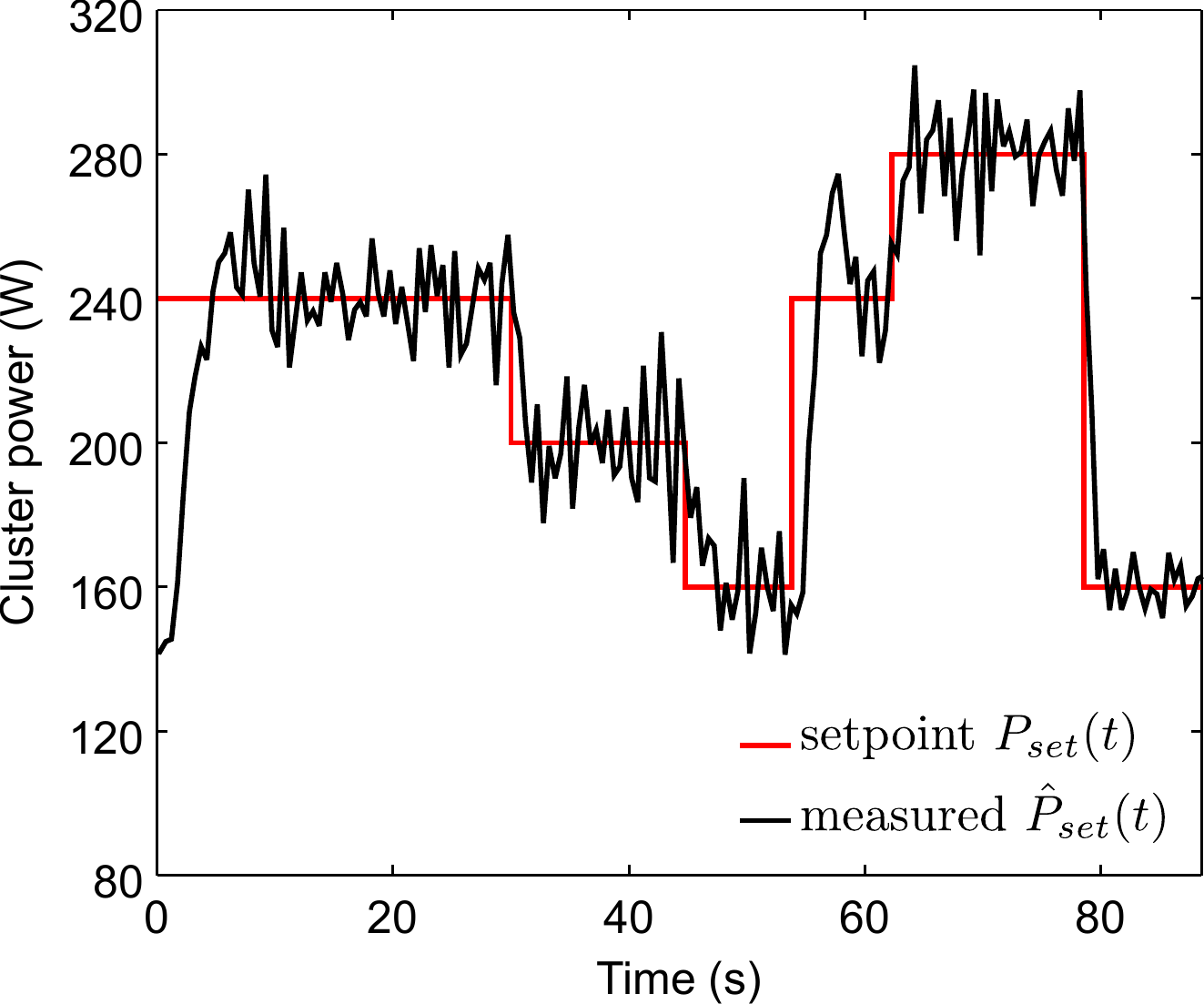}
	\caption{Tracking target with four servers}\label{fig:fourServers}
\end{figure}

In order to track the $P_{set}(t)$ calculated from \cref{eq:powerModel}, the controller on server $i$ simply integrates the received tracking error $e(t)$, and uses a standard anti-windup procedure to produce a signal $\tau_i(t) \in [0,1]$, which represents the fraction of the time that the CPU on server $i$ is active.

We use the ``userspace idle injection'' interface described in \cref{sec:interfaces} (essentially, using Linux signals to duty cycle the workload) to adjust $\tau_i$ whenever the servers receive an updated error signal measurement $e(t)$ (which is updated every $100 \text{ ms}$ in our setup, as noted earlier). Adjustments to $\tau_i$ change server $i$'s power consumption in a software and hardware-dependent manner, indicated by $F_i(\tau)$ in \cref{fig:controlScheme}.

The power tracking experiment shown in \cref{fig:fourServers} shows a maximum settling time of around $3 \text{ s}$ and a root mean square tracking error (see \cref{eq:rmse}) of around $40 \text{ W}$ for the whole cluster. Calculating the PJM-defined \cite{PJMbalancingOps} ``precision score,'' we get a value of $0.86$ for this interval, which is well above the minimum value of $0.75$. It should be noted that this tracking error was achieved with a very basic distributed controller without any communication between nodes -- and might be expected to improve with more advanced control.

\FloatBarrier
\section{Server power model, ramp rate, and dynamic range}\label{sec:suitability}

In the power tracking experiment of \cref{sec:proofOfConcept}, our controller made no assumptions on the control plant. Datacenter power management literature often assumes that the power consumption of a computer server $s$ follows the power model of \cref{eq:powerModel2} with $\tau \in [0,1]$ being percent CPU time \cite{2014GreenEnergyAwarePowerManagement}, and $K_i$, $I_i$ constants.

\begin{equation}\label{eq:powerModel2}
F_{i}(\tau) = K_i \tau + I_i
\end{equation}

\subsection{Accuracy of power model experiment}
We designed an experiment to test the accuracy of the model of \cref{eq:powerModel2} for our servers, under the ``userspace idle injection'' CPU-throttling interface we used in \cref{sec:proofOfConcept}. For this purpose, we first measured the idle power $I_i$, and the dynamic range $K_i$ for each of our four servers. After this, we ran the Linux stress program for $90 \text{ s}$ for $100$ evenly-spaced increments of $\tau$ ranging from $0$ to $1$, recording power measurements in $100 \text{ ms}$ block averages. For each value of $\tau$, we compared $F_{i}(\tau)$ (the server power predicted by \cref{eq:powerModel2}) with $\hat{F}_i(t,\tau)$, the measured server power consumption at time $t$. Using this information, we found the root mean square error (RMSE) according to \cref{eq:rmse} with $T = 100 \text{ ms}$ and $N = 900$.

\begin{equation}\label{eq:rmse}
\text{RMSE}(\tau) = \sqrt{\frac{1}{N}\sum_{n=0}^{N-1}|\hat{F}_{i}(n T, \tau) - F_{i}(\tau)|^2}
\end{equation}

\begin{figure}
	\centering
	\includegraphics[width=\mywidthA]{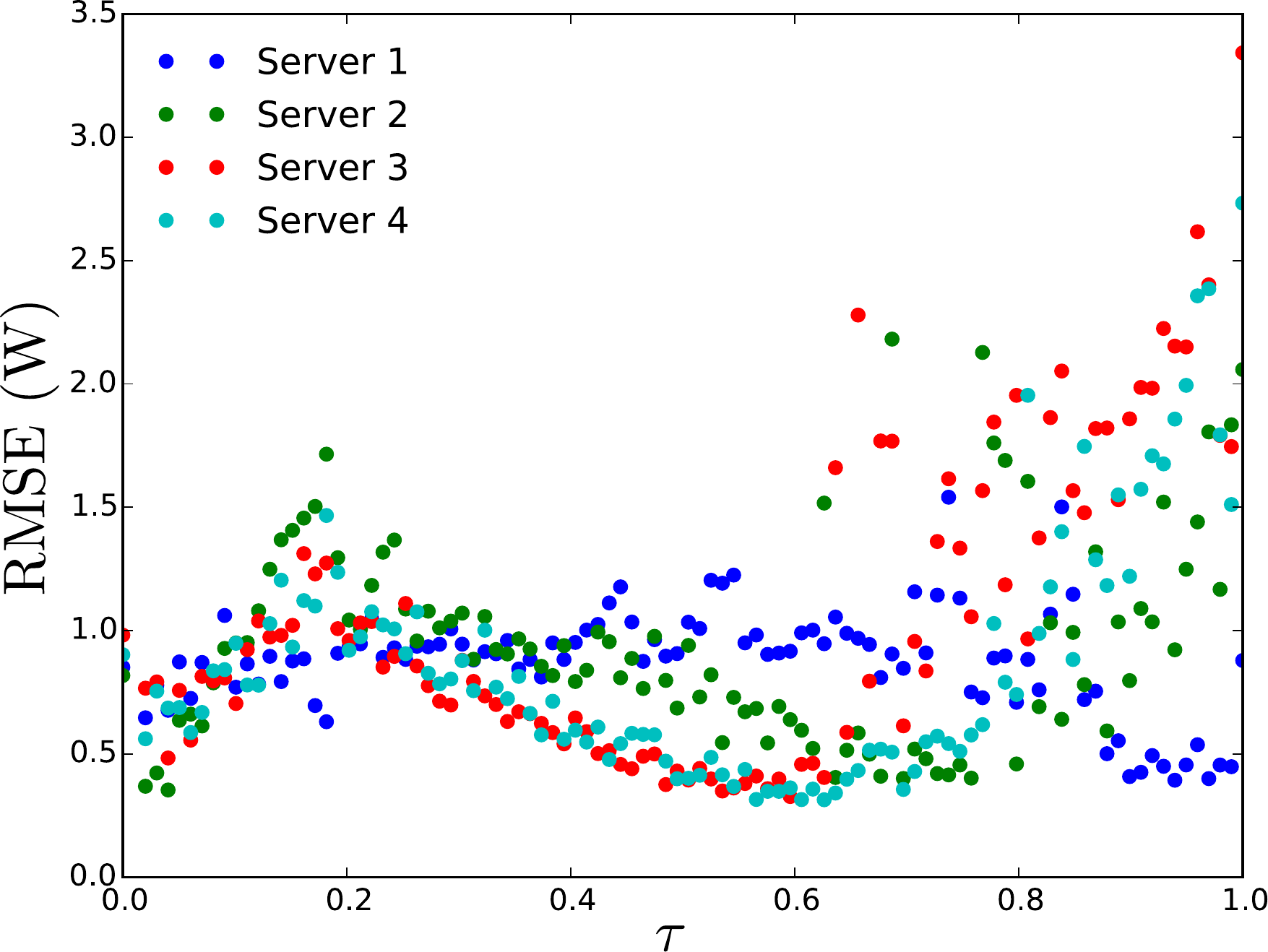}
	\caption{Minimum tracking error in cluster}\label{fig:trackingError}
\end{figure}


Since RMSE compares the measured power with the predicted power at every sample, it gives a more conservative measure of the accuracy of \cref{eq:powerModel2}, than simply comparing $F_i(\tau)$ with the average measured power $\hat{F}_i(\tau)$. \cref{fig:trackingError} plots RMSE versus $\tau$ and shows that the maximum RMSE is around $3 \text{ W}$, meaning that \cref{eq:powerModel2} is a reasonably accurate model for the power consumption of a server using the user-space idle injection method to control the CPU duty cycle $\tau$. 

To put these measurements into context, we can calculate the worst-case PJM ``precision score,'' using \cref{eq:powerModel2} as the setpoint, to see whether a datacenter could feasibly use an open-loop controller based on \cref{eq:powerModel2}. In our measurements, the calculated minimum value of $0.95$ was actually higher than the score we got for our closed-loop controller in part because of the setpoint fluctuations in the previous section, and in part because the simple controller is not tuned to avoid overshoot. An open loop controller has obvious drawbacks, but it is encouraging to note that it is at least in theory feasible.

\subsection{Ramp rate and dynamic range experiment}
\begin{figure}
	\centering
	\includegraphics[width=\mywidthA]{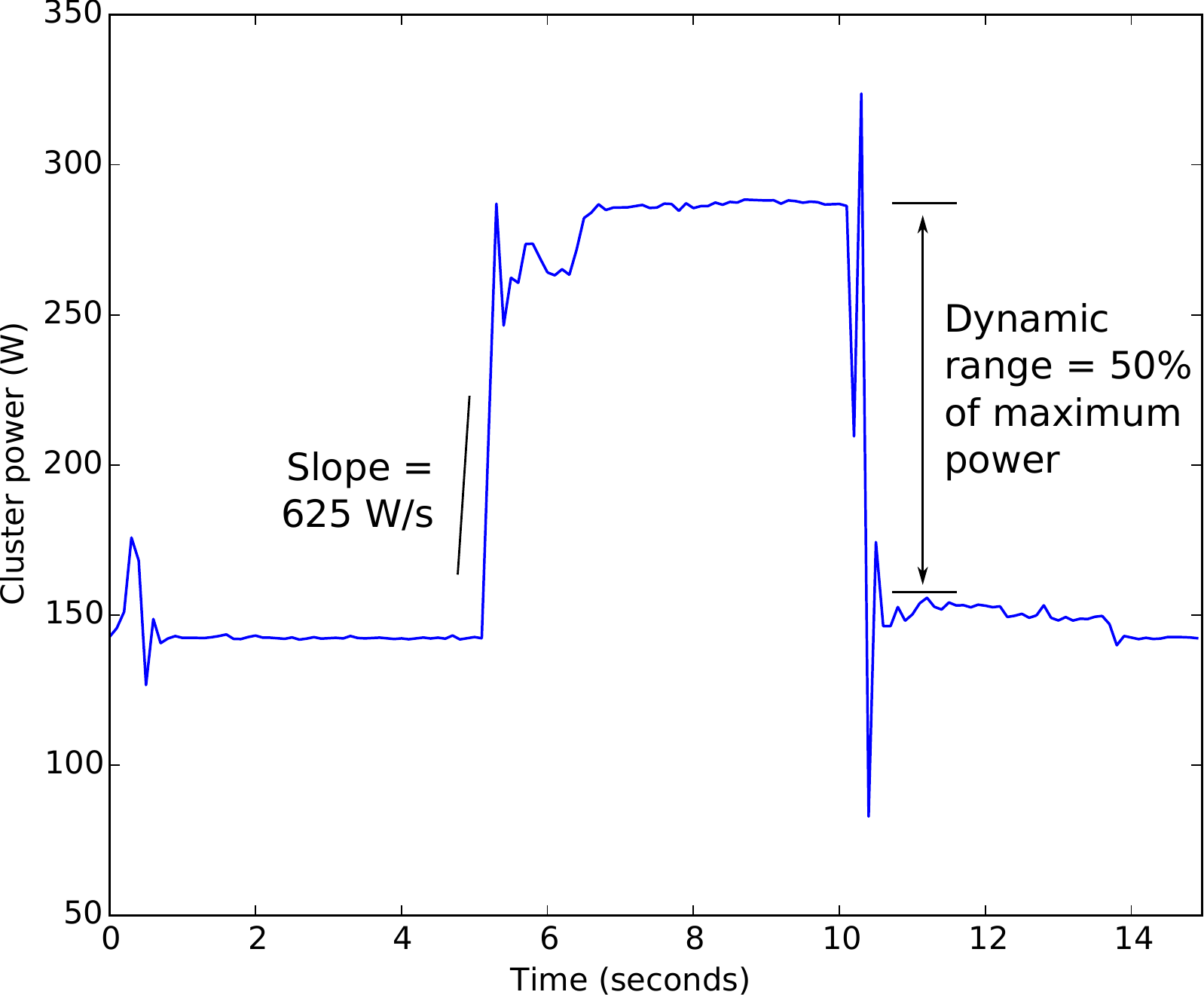}
	\caption{Fast ramp rate of cluster}\label{fig:rampRate}
\end{figure}

Our next experiment is designed to measure the maximum possible ramp rate of the cluster. Our strategy in this experiment was to signal each server in the cluster {\it at exactly the same instant} to transition from an idle CPU state to full load, thereby driving the entire cluster from the lowest to the highest power consumption state in as short a time interval as possible.

In order to achieve this synchronized transition, we wrote a simple remote procedure call (RPC) software to run on each server. This software responded to IP multicast requests to stop and start a computationally-intensive workload. Specifically, we chose the Linux stress \cite{stressProgram} workload generator (performs repeated square root operations) for this experiment, but similar results were obtained with a variety of other computationally-intensive workloads (k-means clustering, financial markets simulation, and video transcoding are a few examples). As \cref{fig:rampRate} shows, it is possible to obtain a fast power ramp rate of about 
$625 \text{ W}/\text{s}$, over a dynamic range of around $145 \text{ W}$, or $50\%$ of the maximum power. To put this into context, consider scaling these results up to a $10 \text{ MW}$, $1.5 \text{ PUE}$ datacenter with a fourth of its servers designated to participate in demand response. At this scale, the datacenter could absorb almost a megawatt, at a rate of up to $4 \text{ MW}/\text{s}$.

\FloatBarrier
\section{Power modulating interfaces}\label{sec:interfaces}
In the experiments described in \cref{sec:proofOfConcept,sec:suitability}, we used the userspace idle injection interface for throttling the CPU time. However, there are a number of different software methods for throttling CPU time, each with unique features and availability. Furthermore, each of these methods produces different statistics for the measured power $\hat{F}_i(\tau)$. As it turns out, some methods are more suited to power control than others. For example, the userspace idle injection interface we used earlier is more deterministic and more closely matches \cref{eq:powerModel2} than the other methods we investigated. Some previous work such as \cite{powerAwareMetering} have produced good experimental measurements in the context of power-aware metering, and previous studies like \cite{workloadDependentDynamicPower} have made contributions to the theoretical side of modeling server power. We complement these studies with a detailed experimental investigation that focuses on the specific software interfaces used to modulate the server power consumption.

We used measured power samples $\hat{F}_i(\tau)$, across the full range of $\tau \in [0,1]$ to estimate some power statistics for each interface. Furthermore, to gain insight into {\it why} the power statistics look as they do, we also measured the percentage of time spent by the processor in each of its active and idle states (in the following, we refer to this information as ``state residency'') using hardware counters in a modified version of Intel's turbostat \cite{turbostat} utility.

\subsection{Experiment description}

In the experiments described in \cref{sec:proofOfConcept,sec:suitability}, we used a ``userspace idle injection" interface for controlling the CPU time. We now justify this choice of interface by presenting a detailed experimental comparison with a number of alternative software methods for controlling CPU utilization. Each of these methods have a different power characteristic $F_i(\tau)$. It turns out that the userspace idle injection interface we used earlier is more deterministic and more closely matches \cref{eq:powerModel2} compared to the other methods which makes it more suitable for our power tracking application.

Throughout the experiment, we recorded both the server power consumption (averaged in $500 \text{ ms}$ blocks) and processor state residency information (again, collected in $500 \text{ ms}$ blocks), and grouped those samples by $\tau$. We removed outliers from each group using the median absolute deviation method \cite{medianAbsoluteDeviation}. And, for each sample group, we recorded the mean and a conservative estimate\footnote{The observed range after outlier rejection, plus $3$ standard deviations.} of the sample range. We performed this experiment for each of the four servers in our cluster, and present results from one server for space considerations and because the power and state residency statistics were nearly identical across the four servers.

If the model of \cref{eq:powerModel2} were correct, plotting the mean of measured power $\hat{F}_{i}(\tau)$ against $\tau$ should produce a straight line with slope $K_i$ (the dynamic range of the server), and y-intercept $I_i$ (the idle power). And for many interfaces, this is close to what we observe, as is shown by the near-linear slope of \cref{fig:cgroups}.

\subsection{Linux cgroups interface}
\begin{figure}
	\centering
	\includegraphics[width=\mywidthB]{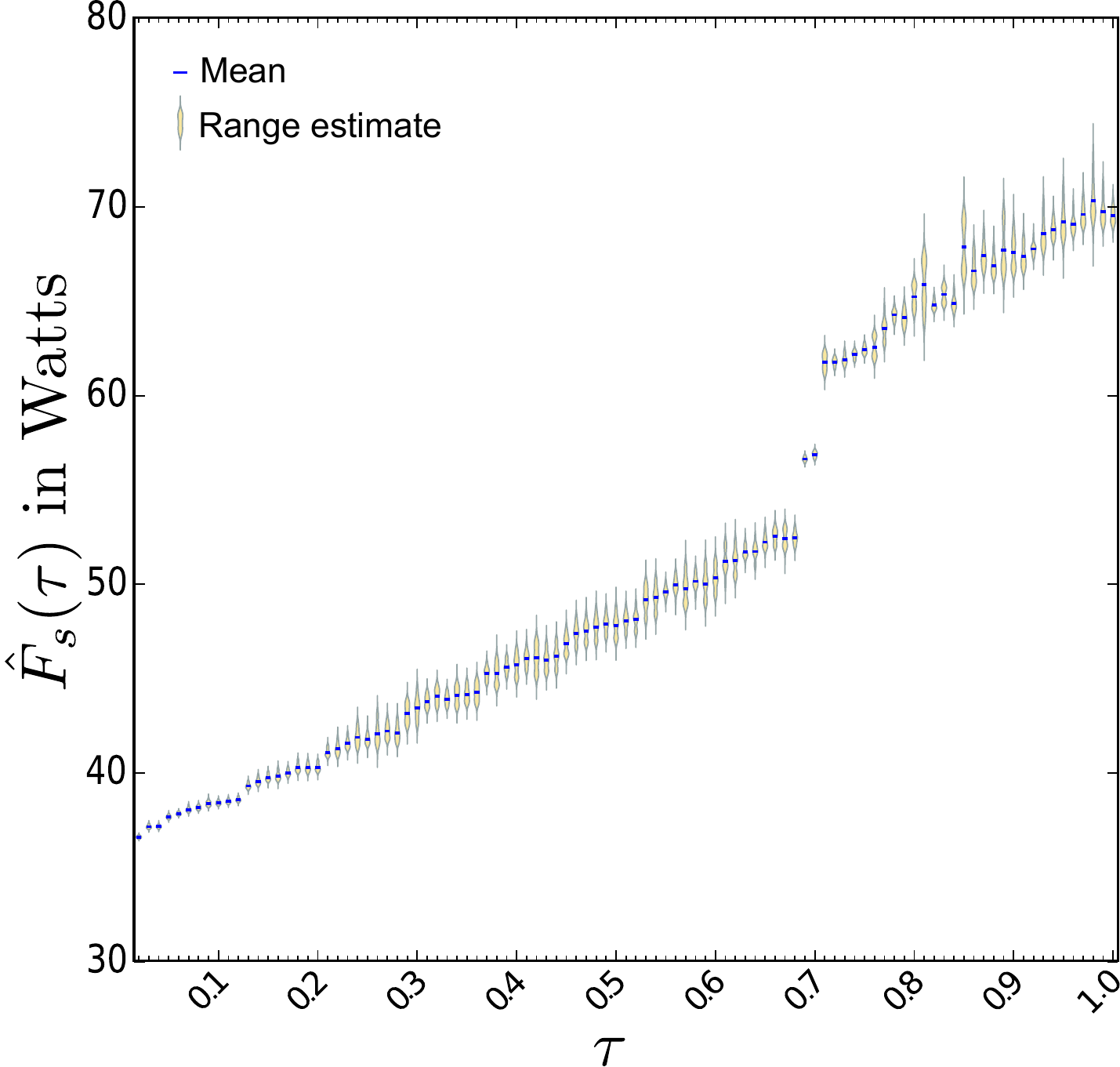}
	\caption{Power profile of server 4 under cgroups interface}
	\label{fig:cgroups}
\end{figure}

\begin{figure*}
	\centering
	\includegraphics[width=\mywidthC]{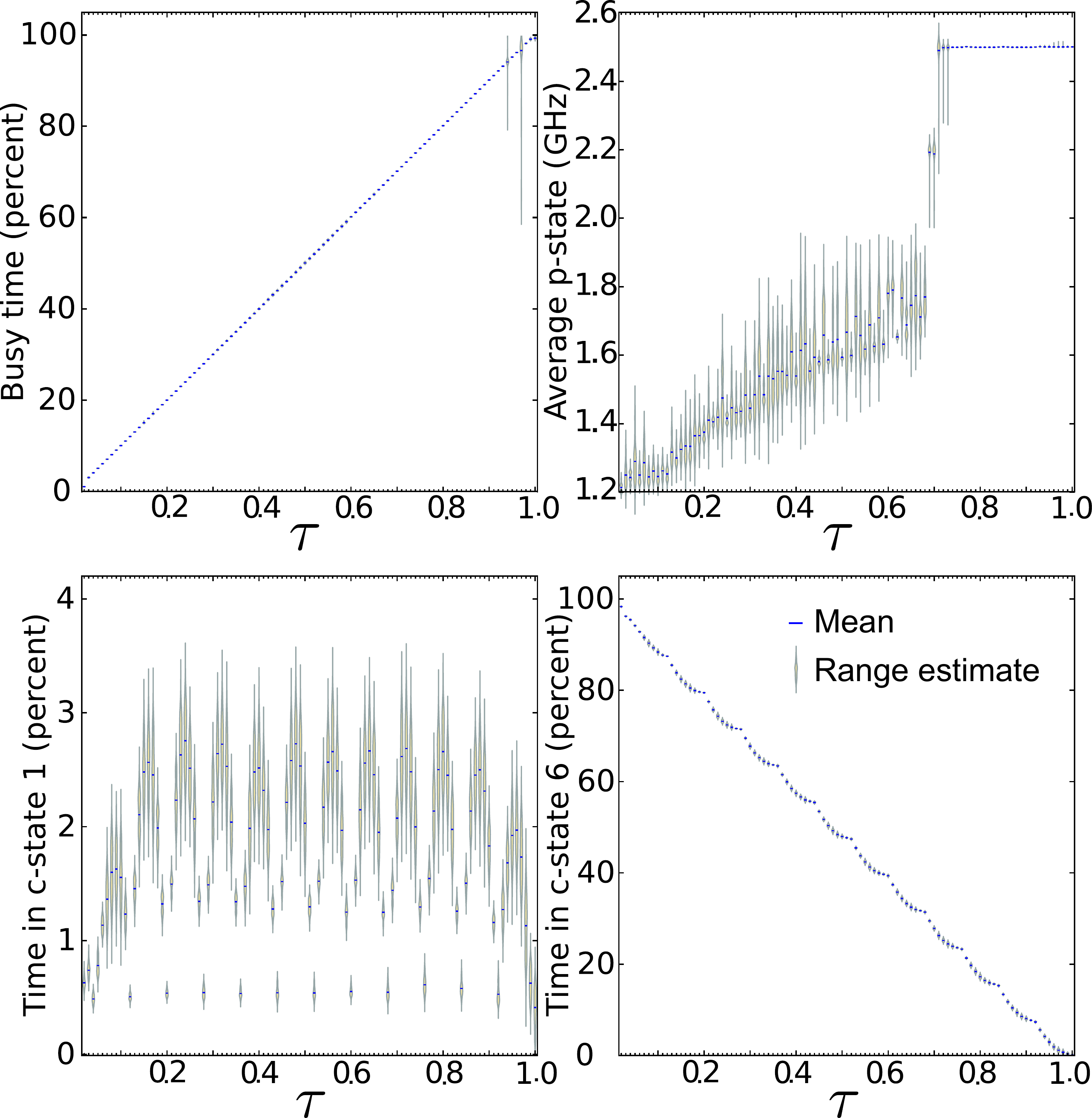}
	\caption{Residency distribution of server 4 under cgroups interface}
	\label{fig:cgroupsResidency}
\end{figure*}

\begin{figure}
	\centering
	\includegraphics[width=\mywidthB]{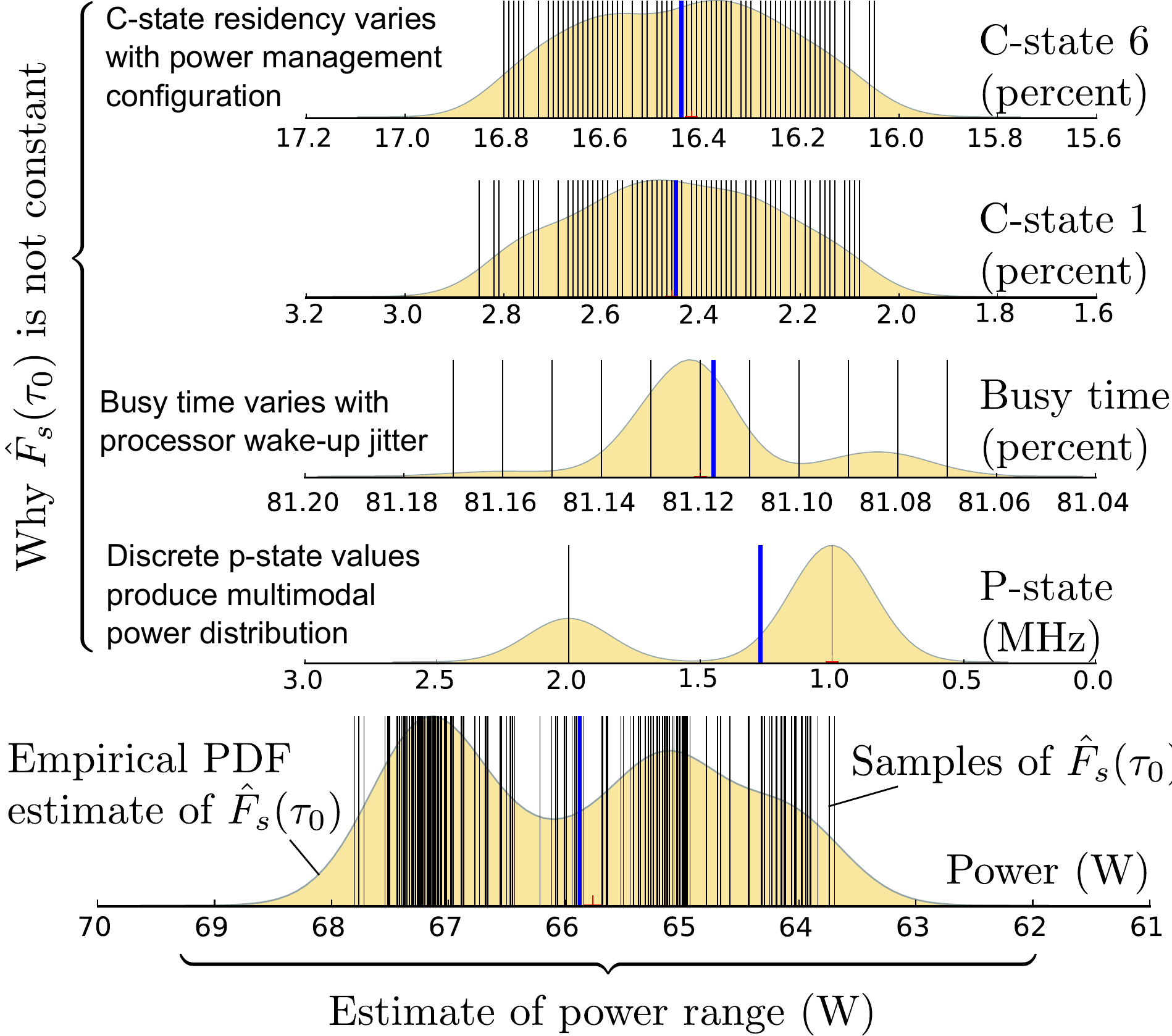}
	\caption{Detailed look at power and processor state samples for $\tau_0 = 0.81$}
	\label{fig:beanPlot}
\end{figure}
The ``user-space idle injection'' interface that we used in our previous experiments is a custom implementation of a standard and more widely used CPU-capping framework called cgroups. 
We thus begin our discussion of the different power modulating interfaces by looking in detail at the Linux cgroup interface. Like many other interfaces, cgroups uses a general technique called idle cycle injection (ICI) to rapidly pause and resume processes in order to limit the average workload ``seen'' by the processor. Unlike direct dynamic voltage and frequency scaling (DVFS) methods which control processor state without modifying the computational workload, ICI can be used to access the full dynamic range of server power. Because ICI does not directly control processor state, the power consumption at a particular $\tau$ may have a larger variance than a DVFS interface would produce. Because of this variance, ICI has been traditionally used in long-term average power control applications like thermal management \cite{Kumar2011, Sironi2013} rather than direct power control. However, our results show that ICI can be quite effective at controlling server power consumption, even on short time scales.

To understand why $F_i(\tau)$ shows nondeterministic behavior, consider \cref{fig:beanPlot}, which gives a detailed look at the observed processor behavior and measured power consumption for the cgroups interface at $\tau = 0.81$. Cgroup's default 
scheduler allows setting a CPU time cap within a periodic interval (set to $100 \text{ ms}$ in our case). Once the workload exceeds its allotted time cap, the operating system pauses the workload processes, and power management features 
automatically transitions the server to one of its idle states. Not only is there some variation to which idle state gets entered, but there is some jitter in the processor's wake-up cycle -- both of which work to producing a non-constant power consumption, even at a constant $\tau$.

\cref{fig:cgroupsResidency} gives a look at the processor behavior under the cgroups interface, across the full range of $\tau$. At the top left, the ``Busy time'' graph has a unit slope, and a small estimated range for each $\tau$ -- showing that the cgroups interface is effective at enforcing its CPU time cap. At the top right, the ``Average p-state'' graph gives a reason for the nonlinearity around $\tau = 0.7$ in the power graph of \cref{fig:cgroups}. At this $\tau$, the power management system stops using the lower-power active modes of the processor, and starts running the processor at maximum speed whenever it is active. At the lower left and right, the ``Time in c-state'' graphs show that the power management system under the cgroups interface prefers to let the processor idle in its lowest-power sleep mode (c-state 6).

Compared to the userspace idle injection interface we used in our previous experiments, one of the main advantages of the cgroups interface is that it has been built in to the Linux kernel since 2008, and is already in use for managing resources in several cluster computing environments (for example, YARN's LinuxContainerExecutor and MESOS's default containerizer). Furthermore, cgroups is designed to work on arbitrary groups of processes and therefore can throttle some groups of processes independently from others. The primary disadvantage is that this method is tied to Linux, while other methods may apply across many different operating system kernels.

\subsection{Other idle cycle injection interfaces}
\subsubsection{Userspace idle injection}
Inspired by the Linux cpulimit program, 
we designed a custom userspace tool to perform ICI using the Linux SIGSTOP and SIGCONT signals. We ran the same experiment as for the cgroups interface. This tool does the same thing as the cgroups interface, except that our tool operates in user space and throttles processes in a more synchronized manner. As shown in \cref{fig:residencyMeans}, this causes the power management system to essentially ``duty cycle'' the processor -- toggling between sleep mode and the highest active state. Because the power management system decides to stop using low power active processor modes around $\tau = 0.2$ rather than cgroup's $\tau = 0.7$, the $\hat{F}_i(\tau)$ for the userspace interface more closely matches $F_i(\tau)$ predicted by \cref{eq:powerModel2}.

The power profile of this tool is the least variable and most linear of all ICI interfaces we tested. The advantage of such a tool is that it can be run in the background of any POSIX operating system, without accessing operating system or hypervisor power management policies. The disadvantage is that it is a custom tool, which only offers marginal improvements over the well-maintained and more full-featured cgroups interface.

\subsubsection{Hypervisor CPU capping}

Various datacenter power management papers such as \cite{ChengWorkload2014} have made use of hypervisor-based CPU throttling techniques. The Xen Project \cite{BarhamXen2003} provides a widely-deployed open source hypervisor, whose sched-credit framework uses ICI to arbitrate processor access among virtual machines (VMs).
The current implementation uses a priority queue and an accounting thread to ensure that the virtual CPUs assigned to the VM do not exceed their CPU use cap for each 30-ms accounting interval.

\begin{figure}[h]
	\centering
	\includegraphics[width=\mywidthB]{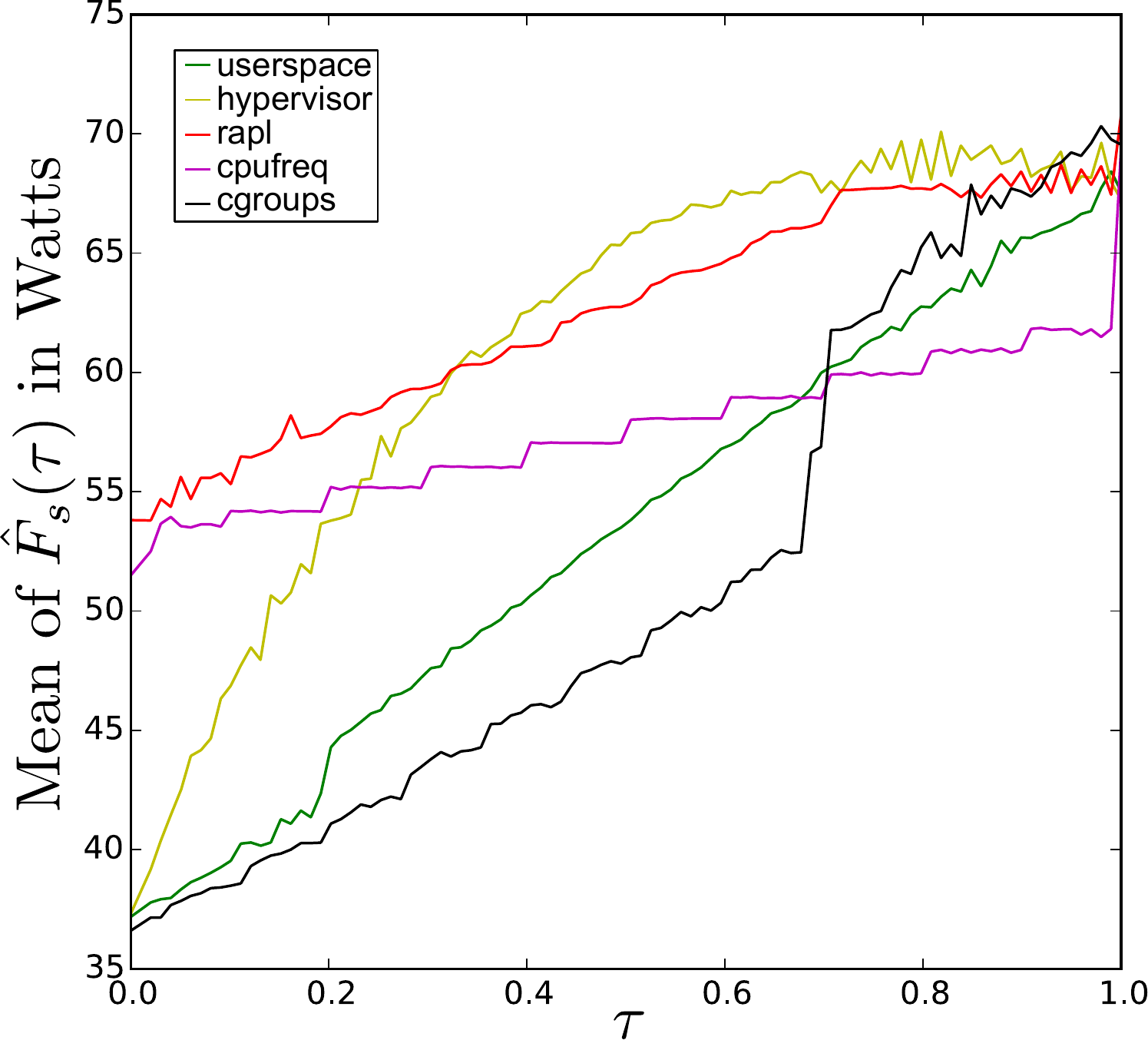}
	\caption{Comparison of mean of $\hat{F}(\tau)$ for server 4 under different power modulating interfaces}
	\label{fig:powerMeans}
\end{figure}

In order to run a workload comparable with the other interfaces, we instantiated a paravirtualized linux guest VM allocated twelve virtual CPUs (two for each hyperthreaded physical core), and ran the same repeated-square-root workload as for the cgroups interface. Compared to the cgroups interface, the power variance at each interface setpoint is higher and \cref{fig:powerMeans} shows that the power profile achieves the full dynamic range but is highly nonlinear. Similar to the cgroups interface, the hypervisor-based interface has the advantage of being integrated into existing virtualized environments and the disadvantage of being tied to those environments. Xen's sched-credit interface has the further inconvenience of a highly nonlinear setpoint-to-power characteristic.

\begin{figure}[h]
	\centering
	\includegraphics[width=\mywidthB]{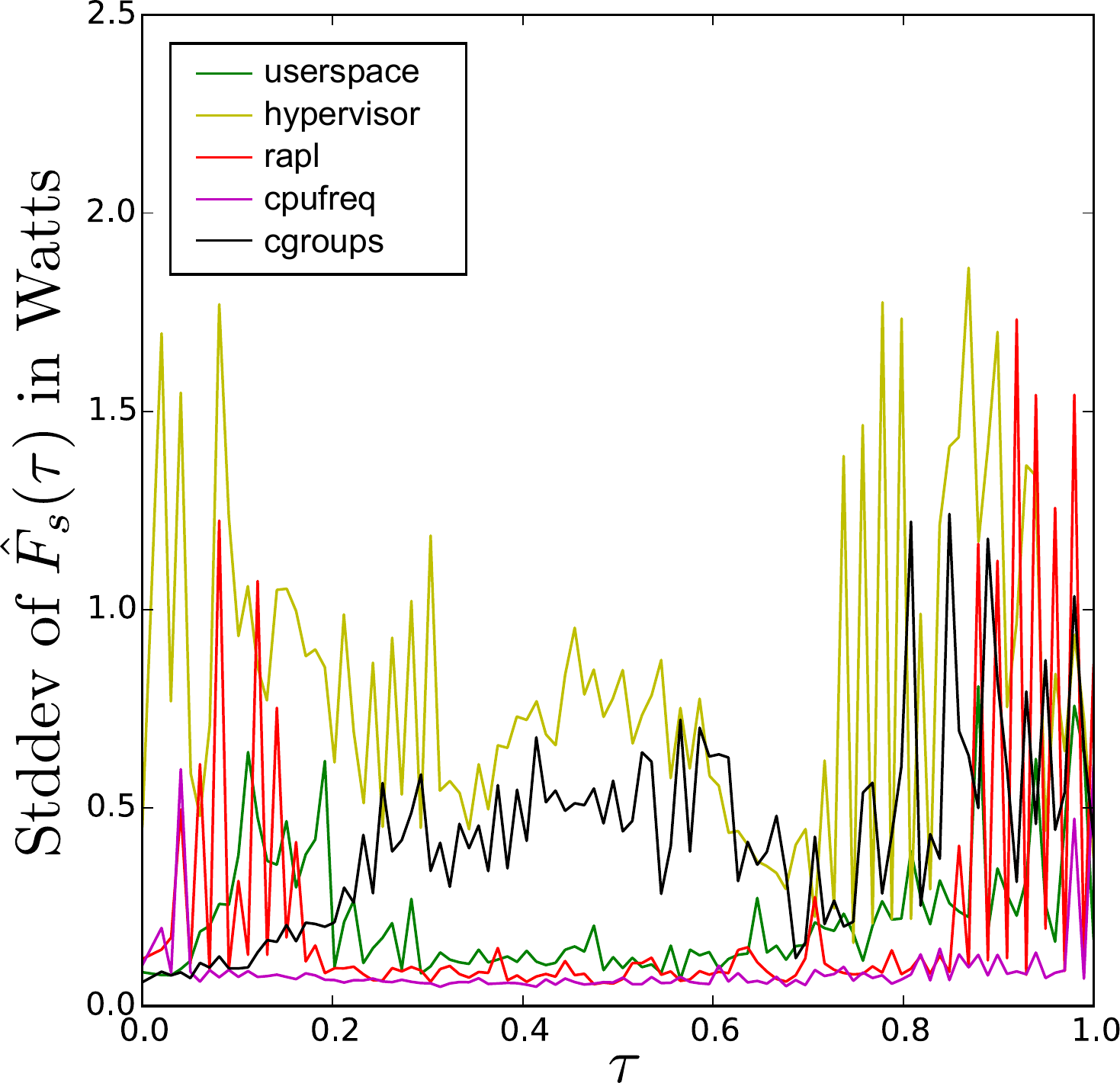}
	\caption{Comparison of standard deviation of $\hat{F}(\tau)$ for server 4 under different power modulating interfaces}
	\label{fig:powerStddevs}
\end{figure}

\subsection{Direct voltage and frequency scaling interfaces}
Rather than throttling workload and letting power management automatically transition processor state, dynamic voltage and frequency scaling (DVFS) interfaces communicate with the processor package directly in order control its operating state. As such, these interfaces offer much tighter control over the server power consumption. However, because they do not cause the processor to enter sleep states, DVFS interfaces do not allow power control over the server's full dynamic range.

\subsubsection{Cpufreq driver}
\begin{figure*}
	\centering
	\includegraphics[width=\mywidthC]{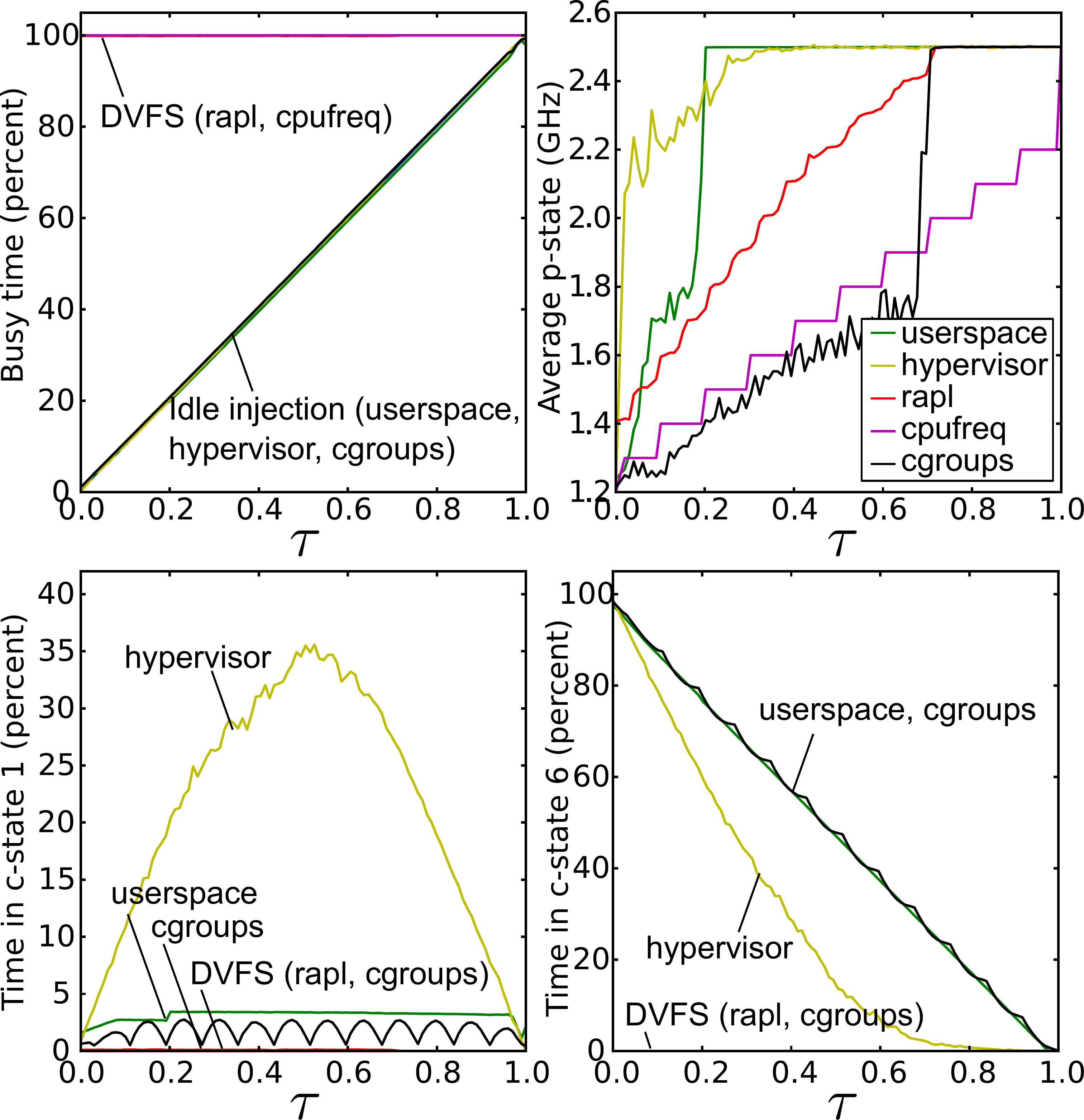}
	\caption{Comparison of mean state residency profile for server 4 under different power modulating interfaces}
	\label{fig:residencyMeans}
\end{figure*}

For example, several papers like \cite{LiTAPA2011,Karpowicz2013} have successfully applied the userspace governor of the Linux acpi-cpufreq legacy driver to implement energy-aware DVFS on older processors. However, it is worth noting that most modern Intel processors only allow DVFS ``hinting'' rather than direct control. 
The more modern intel\_pstate driver offloads power consumption to hardware and does not include a userspace governor. This technique has the advantage of a very low power variance at each setpoint, but only about half of the dynamic range is controllable. Further, as \cref{fig:residencyMeans} shows, the cpufreq driver only allows twelve distinct voltage-frequency pairs, meaning that the power characteristic of \cref{fig:powerMeans} has only twelve unique levels.

\subsubsection{Proprietary On-chip Power-Limiting}
The next DVFS technique is an improvement to the quantized cpufreq interface. RAPL precisely limits the processor package and the memory power consumption using a proprietary hardware mechanism, but the observed state residency (see \cref{fig:powerMeans}) indicates that a DVFS technique is used. Certain Intel processors support a model-specific register (MSR) interface called Running Average Power Limit (RAPL) \cite{RAPL}. Papers like \cite{Lo2014} use the RAPL API  to shape the power consumption of servers running transactional workloads to achieve high energy efficiency with minimal service-level objective violations. This technique again has very low power variance at each setpoint and covers almost the same dynamic range as the cpufreq interface. But, it is not limited to discrete levels of power consumption, as in the cpufreq interface.


\section{Discussion} \label{sec:discussion}

The measurements in \cref{sec:interfaces} showed that server power can be controlled with fast ramp rate, across a wide dynamic range, and with a high level of precision using several different power modulating software interfaces. 


Out of all the power-modulating interfaces discussed in \cref{sec:interfaces}, the Linux cgroups interface stands out because of its low power variance, ubiquitous availability (almost all Linux servers have this feature), and highly flexible nature (allows different process groups to be managed separately). However, it is important to consider the specific needs of that datacenter when selecting a power modulating interface for servers in a realtime datacenter demand response system. For example, Xen's sched-credit interface may be more applicable in a highly virtualized environment, or the RAPL interface may be more appropriate if precise controllability is valued over dynamic range. Even direct DVFS or custom software solutions may offer the best tradeoff for certain applications.

Furthermore, our experiments have shown that the linear power model of \cref{eq:powerModel2} commonly used in datacenter power management literature is not always valid, depending on the interface used. Server power $F_i(\tau)$, with CPU time $\tau$ controlled by Xen's sched-credit interface, for example, exhibits highly nonlinear dependence on $\tau$. The results of our comparison of the software power control interfaces can be summarized as follows.

\begin{itemize}
	\item The linear model of \cref{eq:powerModel2} relating server power to CPU time holds reasonably well for the Linux cgroups, userspace idle injection, and RAPL interfaces, but is not a good model for the Xen sched-credit interface or the cpufreq driver interface.
	\item Idle Cycle Injection is very effective in providing a wide dynamic range but suffers from increased power variance.
	\item Direct Dynamic Voltage and Frequency Scaling is effective in accurate power control with low variance, but offers a limited dynamic range.
\end{itemize}



\section{Conclusion} \label{sec:concl}
In this paper, we have provided evidence that datacenters are attractive controllable loads. And, as a first step toward understanding the details of implementing a controller for datacenter power shaping we have compared and contrasted several different classes of cpu-utilization capping software available to rack servers. Custom software used in this paper is open source and can be downloaded from \cite{josiahServer}. Our power tracking experiment showed that a very simple controller can exceed industry power tracking specifications. Further measurements showed a dynamic range of around $50\%$ of the maximum power, and practically no upper limit on the power ramp rate.

An immediate next step in experimental measurement to would be to determine other demand response metrics like power factor and total harmonic distortion when servers are tracking a typical realtime power setpoint. While the integral control scheme in \cref{sec:proofOfConcept} ``works,'' there are likely many senses in which it is not optimal. Determining what characteristics are most desirable in a power controller for realtime demand response is an exciting open question. For example, one of our assumptions in the power tracking experiment was a workload which could be parallelized without job dependencies or strong SLA restrictions. Along those lines, distributed control as in \cite{2012OptimalEconomicDispatch,2015DistributedDemandResponse}, but incorporating the effect of SLA and workload structure (for example, MapReduce jobs) seems to be an attractive starting point for future research.


\ifCLASSOPTIONcompsoc
\section*{Acknowledgments}
\else
\section*{Acknowledgment}
\fi

A preliminary version of this paper has been accepted for presentation in the Innovative Smart Grid Technologies 2016 Conference \cite{isgtJosiah}. This material is based upon work supported by the NSF Graduate Research Fellowship Program under Grant Number DGE-1144245. It is also in part supported by US NSF grants EPS-1101284, CAREER ECCS-1150801, CNS-1239509 and CCF-1302456, and ONR grant N00014-13-1-0202.

\bibliographystyle{IEEEtran}
\bibliography{bibliography}

\begin{thebibliography}{10}
\providecommand{\url}[1]{#1}
\csname url@samestyle\endcsname
\providecommand{\newblock}{\relax}
\providecommand{\bibinfo}[2]{#2}
\providecommand{\BIBentrySTDinterwordspacing}{\spaceskip=0pt\relax}
\providecommand{\BIBentryALTinterwordstretchfactor}{4}
\providecommand{\BIBentryALTinterwordspacing}{\spaceskip=\fontdimen2\font plus
\BIBentryALTinterwordstretchfactor\fontdimen3\font minus
  \fontdimen4\font\relax}
\providecommand{\BIBforeignlanguage}[2]{{%
\expandafter\ifx\csname l@#1\endcsname\relax
\typeout{** WARNING: IEEEtran.bst: No hyphenation pattern has been}%
\typeout{** loaded for the language `#1'. Using the pattern for}%
\typeout{** the default language instead.}%
\else
\language=\csname l@#1\endcsname
\fi
#2}}
\providecommand{\BIBdecl}{\relax}
\BIBdecl

\bibitem{koomey2011growth}
J.~Koomey, ``Growth in data center electricity use 2005 to 2010,'' Analytics
  Press, for The New York Times, Report, 2011.

\bibitem{largestDatacenters}
\BIBentryALTinterwordspacing
{Rich Miller}, ``The world's largest data centers.'' [Online]. Available:
  \url{http://www.datacenterknowledge.com/special-report-the-worlds-largest-data-centers/}
\BIBentrySTDinterwordspacing

\bibitem{powerFluctuations}
\BIBentryALTinterwordspacing
{Patrick Donovan from Schneider Electric}. An overlooked problem: Dynamic power
  variations. [Online]. Available:
  \url{http://www.datacenterknowledge.com/archives/2013/06/21/a-commonly-overlooked-problem-dynamic-power-variations-in-data-centers-and-network-rooms/}
\BIBentrySTDinterwordspacing

\bibitem{2014DemandResponseSurvey}
P.~Siano, ``Demand response and smart grids: A survey,'' \emph{Renewable and
  Sustainable Energy Reviews}, vol.~30, pp. 461 -- 478, 2014.

\bibitem{2012DatacenterDRFieldStudies}
G.~Ghatikar, V.~Ganti, N.~Matson, and M.~A. Piette, ``Demand response
  opportunities and enabling technologies for data centers: Findings from field
  studies,'' 2012.

\bibitem{2013DatacenterCLR}
R.~Wang, N.~Kandasamy, C.~Nwankpa, and D.~R. Kaeli, ``Datacenters as
  controllable load resources in the electricity market,'' in \emph{Proceedings
  of the 2013 IEEE 33rd International Conference on Distributed Computing
  Systems}, ser. ICDCS '13.\hskip 1em plus 0.5em minus 0.4em\relax Washington,
  DC, USA: IEEE Computer Society, 2013, pp. 176--185.

\bibitem{2013RealtimePowerControlOfDatacenters}
H.~Chen, A.~K. Coskun, and M.~C. Caramanis, ``Real-time power control of data
  centers for providing regulation service,'' in \emph{2013 IEEE 52nd Annual
  Conference on Decision and Control}, Dec 2013, pp. 4314--4321.

\bibitem{2014DatacenterDemandResponse}
A.~Wierman, Z.~Liu, I.~Liu, and H.~Mohsenian-Rad, ``Opportunities and
  challenges for data center demand response,'' in \emph{Green Computing
  Conference (IGCC), 2014 International}, Nov 2014, pp. 1--10.

\bibitem{2014DatacenterDynamicPricing}
M.~Ghasemi-Gol, Y.~Wang, and M.~Pedram, ``An optimization framework for data
  centers to minimize electric bill under day-ahead dynamic energy prices while
  providing regulation services,'' in \emph{Green Computing Conference (IGCC),
  2014 International}, Nov 2014, pp. 1--9.

\bibitem{2014DatacenterGridLoadStabilizer}
H.~Chen, M.~C. Caramanis, and A.~K. Coskun, ``The data center as a grid load
  stabilizer,'' in \emph{2014 19th Asia and South Pacific Design Automation
  Conference (ASP-DAC)}, Jan 2014, pp. 105--112.

\bibitem{2014DatacentersAndElectricVehicles}
S.~Li, M.~Brocanelli, W.~Zhang, and X.~Wang, ``Integrated power management of
  data centers and electric vehicles for energy and regulation market
  participation,'' \emph{IEEE Transactions on Smart Grid}, vol.~5, no.~5, pp.
  2283--2294, Sept 2014.

\bibitem{2014PricingDatacenterDemandResponse}
Z.~Liu, I.~Liu, S.~Low, and A.~Wierman, ``Pricing data center demand
  response,'' in \emph{The 2014 ACM International Conference on Measurement and
  Modeling of Computer Systems}, ser. SIGMETRICS '14.\hskip 1em plus 0.5em
  minus 0.4em\relax New York, NY, USA: ACM, 2014, pp. 111--123.

\bibitem{2015DatacenterOptimalRegulationService}
H.~Chen, B.~Zhang, M.~C. Caramanis, and A.~K. Coskun, ``Data center optimal
  regulation service reserve provision with explicit modeling of quality of
  service dynamics,'' in \emph{2015 54th IEEE Conference on Decision and
  Control}, Dec 2015, pp. 7207--7213.

\bibitem{1915MultiplexCostAndRateSystem}
O.~B. Goldman, ``The multiplex cost and rate system,'' \emph{American Institute
  of Electrical Engineers, Proceedings of the}, vol.~34, no.~5, pp. 941--957,
  May 1915.

\bibitem{1987ShortTermLoadForecasting}
G.~Gross and F.~D. Galiana, ``Short-term load forecasting,'' \emph{Proceedings
  of the IEEE}, vol.~75, no.~12, pp. 1558--1573, Dec 1987.

\bibitem{porter2012survey}
{K. Porter} and {J. Rogers}, ``Survey of variable generation forecasting in the
  west,'' National Renewable Energy Laboratory, Subcontract Report
  {NREL}/{SR}-5500-54457, {NREL Technical Monitor: Dr. Kirsten Orwig}.

\bibitem{2013PowerSystemSchedulingWithRenewables}
A.~Hooshmand, J.~Mohammadpour, H.~Malki, and H.~Daneshi, ``Power system dynamic
  scheduling with high penetration of renewable sources,'' in \emph{2013
  American Control Conference}, June 2013, pp. 5827--5832.

\bibitem{2007IntermittentGeneration}
A.~D. Lamont, ``Assessing the long-term system value of intermittent electric
  generation technologies,'' \emph{Energy Economics}, vol.~30, no.~3, pp. 1208
  -- 1231, 2008.

\bibitem{2011DemandResponseSmartLoads}
P.~Palensky and D.~Dietrich, ``Demand side management: Demand response,
  intelligent energy systems, and smart loads,'' \emph{IEEE Transactions on
  Industrial Informatics}, vol.~7, 2011.

\bibitem{misoACE}
\BIBentryALTinterwordspacing
{Midcontinent Independent System Operator}, ``{ACE} chart.'' [Online].
  Available:
  \url{https://www.misoenergy.org/MARKETSOPERATIONS/REALTIMEMARKETDATA/Pages/ACEChart.aspx}
\BIBentrySTDinterwordspacing

\bibitem{primarycontrol}
F.~R. S.~D. Team, ``Frequency response standard background document,'' North
  American Electric Reliability Corporation, Reference Information for
  BAL‐003‐1.

\bibitem{2011GreenScheduling}
B.~Aksanli, J.~Venkatesh, L.~Zhang, and T.~Rosing, ``Utilizing green energy
  prediction to schedule mixed batch and service jobs in data centers,'' in
  \emph{Proceedings of the 4th Workshop on Power-Aware Computing and Systems},
  ser. HotPower '11.\hskip 1em plus 0.5em minus 0.4em\relax New York, NY, USA:
  ACM, 2011, pp. 5:1--5:5.

\bibitem{2013ParasolAndGreenSwitch}
I.~n. Goiri, W.~Katsak \emph{et~al.}, ``Parasol and greenswitch: Managing
  datacenters powered by renewable energy,'' in \emph{Proceedings of the
  Eighteenth International Conference on Architectural Support for Programming
  Languages and Operating Systems}, ser. ASPLOS '13.\hskip 1em plus 0.5em minus
  0.4em\relax New York, NY, USA: ACM, 2013, pp. 51--64.

\bibitem{energyAwareResourceAllocation}
X.~Xu, W.~Dou, X.~Zhang, and J.~Chen, ``Enreal: An energy-aware resource
  allocation method for scientific workflow executions in cloud environment,''
  \emph{IEEE Transactions on Cloud Computing}, vol.~4, no.~2, pp. 166--179,
  April 2016.

\bibitem{2014GreenEnergyAwarePowerManagement}
F.~Kong and X.~Liu, ``A survey on green-energy-aware power management for
  datacenters,'' \emph{ACM Comput. Surv.}, vol.~47, no.~2, pp. 30:1--30:38,
  Nov. 2014.

\bibitem{WangCLR2013}
R.~Wang, N.~Kandasamy, C.~Nwankpa, and D.~R. Kaeli, ``Datacenters as
  controllable load resources in the electricity market,'' in \emph{Proceedings
  of the 2013 IEEE 33rd International Conference on Distributed Computing
  Systems}, ser. ICDCS '13.\hskip 1em plus 0.5em minus 0.4em\relax Washington,
  DC, USA: IEEE Computer Society, 2013, pp. 176--185.

\bibitem{ChengWorkload2014}
D.~Cheng, C.~Jiang, and X.~Zhou, ``Heterogeneity-aware workload placement and
  migration in distributed sustainable datacenters,'' in \emph{Parallel and
  Distributed Processing Symposium, 2014 IEEE 28th International}, May 2014,
  pp. 307--316.

\bibitem{PJMbalancingOps}
C.~Pilong, ``Pjm manual 12: Balancing operations,'' PJM Interconnection, Manual
  Revision: 34.

\bibitem{PJMmanual28}
M.~S.~D. Department, ``Pjm manual 28: Operating agreement accounting,'' PJM
  Interconnection, Manual Revision: 73.

\bibitem{koomey2007servers}
J.~Koomey, ``Estimating total power consumption by servers in the {U.S.} and
  the world,'' Analytics Press, Report, 2011.

\bibitem{2004VideoCodingClustersOfWorkstations}
A.~Rodriguez, A.~Gonzalez, and M.~P. Malumbres, ``Performance evaluation of
  parallel mpeg-4 video coding algorithms on clusters of workstations,'' in
  \emph{Parallel Computing in Electrical Engineering, 2004. PARELEC 2004.
  International Conference on}, Sept 2004, pp. 354--357.

\bibitem{amazonTranscoderFAQ}
\BIBentryALTinterwordspacing
{Amazon}, ``Amazon {AWS} elastic transcoder {FAQ}.'' [Online]. Available:
  \url{https://aws.amazon.com/elastictranscoder/faqs/}
\BIBentrySTDinterwordspacing

\bibitem{joeThesis}
J.~E. Hall, ``Distributed control system for demand response by servers,''
  Master's thesis, University of Iowa, 2015.

\bibitem{stressProgram}
\BIBentryALTinterwordspacing
{Amos Waterland}, ``stress {POSIX} workload generator.'' [Online]. Available:
  \url{http://people.seas.harvard.edu/~apw/stress/}
\BIBentrySTDinterwordspacing

\bibitem{powerAwareMetering}
A.~Narayan and S.~Rao, ``Power-aware cloud metering,'' \emph{IEEE Transactions
  on Services Computing}, vol.~7, no.~3, pp. 440--451, July 2014.

\bibitem{workloadDependentDynamicPower}
K.~Li, ``Improving multicore server performance and reducing energy consumption
  by workload dependent dynamic power management,'' \emph{IEEE Transactions on
  Cloud Computing}, vol.~4, no.~2, pp. 122--137, April 2016.

\bibitem{turbostat}
\BIBentryALTinterwordspacing
{Len Brown}, ``Ubuntu 14.04 manpages: turbostat.'' [Online]. Available:
  \url{http://manpages.ubuntu.com/manpages/trusty/man8/turbostat.8.html}
\BIBentrySTDinterwordspacing

\bibitem{medianAbsoluteDeviation}
C.~Leys, C.~Ley \emph{et~al.}, ``Detecting outliers: Do not use standard
  deviation around the mean, use absolute deviation around the median,''
  \emph{Journal of Experimental Social Psychology}, vol.~49, no.~4, pp. 764 --
  766, 2013.

\bibitem{Kumar2011}
P.~Kumar and L.~Thiele, ``Cool shapers: Shaping real-time tasks for improved
  thermal guarantees,'' in \emph{Design Automation Conference (DAC), 2011 48th
  ACM/EDAC/IEEE}, June 2011, pp. 468--473.

\bibitem{Sironi2013}
F.~Sironi, M.~Maggio \emph{et~al.}, ``Thermos: System support for dynamic
  thermal management of chip multi-processors,'' in \emph{Parallel
  Architectures and Compilation Techniques (PACT), 2013 22nd International
  Conference on}, Sept 2013, pp. 41--50.

\bibitem{BarhamXen2003}
P.~Barham, B.~Dragovic \emph{et~al.}, ``Xen and the art of virtualization,'' in
  \emph{Proceedings of the Nineteenth ACM Symposium on Operating Systems
  Principles}, ser. SOSP '03.\hskip 1em plus 0.5em minus 0.4em\relax New York,
  NY, USA: ACM, 2003, pp. 164--177.

\bibitem{LiTAPA2011}
S.~Li, T.~Abdelzaher, and M.~Yuan, ``Tapa: Temperature aware power allocation
  in data center with map-reduce,'' in \emph{Green Computing Conference and
  Workshops (IGCC), 2011 International}, July 2011, pp. 1--8.

\bibitem{Karpowicz2013}
M.~Karpowicz, ``On the design of energy-efficient service rate control
  mechanisms: Cpu frequency control for linux,'' in \emph{Digital
  Communications - Green ICT (TIWDC), 2013 24th Tyrrhenian International
  Workshop on}, Sept 2013, pp. 1--6.

\bibitem{RAPL}
H.~David, E.~Gorbatov \emph{et~al.}, ``Rapl: Memory power estimation and
  capping,'' in \emph{Low-Power Electronics and Design (ISLPED), 2010 ACM/IEEE
  International Symposium on}, Aug 2010, pp. 189--194.

\bibitem{Lo2014}
D.~Lo, L.~Cheng \emph{et~al.}, ``Towards energy proportionality for large-scale
  latency-critical workloads,'' in \emph{Proceeding of the 41st Annual
  International Symposium on Computer Architecuture}, ser. ISCA '14.\hskip 1em
  plus 0.5em minus 0.4em\relax Piscataway, NJ, USA: IEEE Press, 2014, pp.
  301--312.

\bibitem{josiahServer}
\BIBentryALTinterwordspacing
{Josiah McClurg}, ``Serverpower repository.'' [Online]. Available:
  \url{https://github.com/jcmcclurg/serverpower}
\BIBentrySTDinterwordspacing

\bibitem{2012OptimalEconomicDispatch}
R.~Mudumbai, S.~Dasgupta, and B.~Cho, ``Distributed control for optimal
  economic dispatch of a network of heterogeneous power generators,''
  \emph{Power Systems, IEEE Transactions on}, vol.~27, no.~4, pp. 1750--1760,
  Nov 2012.

\bibitem{2015DistributedDemandResponse}
S.~Goguri, J.~Hall, R.~Mudumbai, and S.~Dasgupta, ``A distributed, real-time
  and non-parametric approach to demand response in the smart grid,'' in
  \emph{Information Sciences and Systems (CISS), 2015 49th Annual Conference
  on}, March 2015, pp. 1--5.

\bibitem{isgtJosiah}
J.~McClurg, J.~Hall, and R.~Mudumbai, ``Fast demand response with datacenter
  loads,'' in \emph{Innovative Smart Grid Technologies Conference (ISGT), 2016
  IEEE Power Energy Society}, Sept 2016, (accepted).

\end{thebibliography}

%

\begin{IEEEbiography}[{\includegraphics[width=1in,height=1.25in,clip,keepaspectratio]{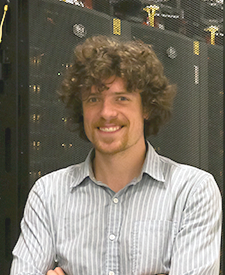}}]{Josiah McClurg}
	is a Ph.D. candidate in Electrical \& Computer Engineering at the University of Iowa. He earned an M.S. in EE (Control Systems focus) from the University of Iowa in 2012, and an M.S. in EE (Power Electronics focus) from the University of Illinois Champaign-Urbana in 2015. His research interests are in power-aware computing, demand response, and other enabling technologies for smart grid.
\end{IEEEbiography}

\begin{IEEEbiography}[{\includegraphics[width=1in,height=1.25in,clip,keepaspectratio]{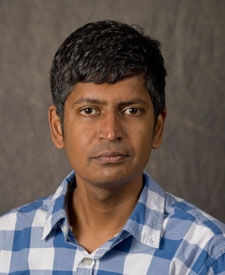}}]{Raghuraman Mudumbai}
	is an Associate Professor in Electrical \& Computer Engineering at the University of Iowa. He obtained his B.Tech in EE from the Indian Institute of Technology, Madras, India in 1998; MSEE from the Polytechnic University, Brooklyn in 2000 and Ph.D in ECE from the University of California Santa Barbara in 2007. He also worked as a Systems Engineer at LM Ericsson Telephone Co between 2001 and 2003. His research interests are in wireless communications and power systems.
\end{IEEEbiography}

\vfill




\end{document}